\documentclass[12pt]{iopart}
\usepackage{iopams}
\newcommand{\be}{\begin{equation*}}
\newcommand{\ee}{\end{equation*}}
\begin{document}

\title[Novel Features of the Energy Momentum Tensor]
{Novel Features of the Energy Momentum Tensor of a Casimir
Apparatus in a Weak Gravitational Field}

\author{Giuseppe Bimonte\ddag\P, Enrico Calloni\ddag\P, 
Giampiero Esposito\P, Luigi Rosa\ddag\P}

\address{\ddag Dipartimento di Scienze Fisiche, Complesso Universitario
di Monte S. Angelo \\
Via Cintia, Edificio 6, 80126 Napoli, Italy}
\address{\P INFN, Sezione di Napoli, Complesso Universitario di Monte S.
Angelo \\
Via Cintia, Edificio 6, 80126 Napoli, Italy}

\abstract The influence of the gravity acceleration on the
regularized energy-momentum tensor of the quantized
electromagnetic field between two plane parallel conducting plates
is derived. A perturbative expansion, to first order in the
constant acceleration parameter, of the Green functions involved
and of the energy-momentum tensor is derived by means of the
covariant geodesic point splitting procedure. The energy-momentum
tensor is covariantly conserved and satisfies the expected
relation between gauge-breaking and ghost parts. 
\endabstract

\section{Introduction}

An important property of quantum electrodynamics is that suitable
differences of zero-point energies of the quantized
electromagnetic field can be made finite and produce measurable
effects such as the tiny attractive force among perfectly
conducting parallel plates known as the Casimir effect
\cite{Bord01}. This is a remarkable quantum mechanical effect that
makes itself manifest on a macroscopic scale. For perfect
reflectors and metals the Casimir force can be attractive or
repulsive, depending on the geometry of the cavity, whereas for
dielectrics in the weak-reflector approximation it is always
attractive, independently of the geometry \cite{Bart01}. The
Casimir effect can be studied within the framework of boundary
effects in quantum field theory, combined with zeta-function
regularization or Green-function methods, or in more physical
terms, i.e. on considering van der Waals forces \cite{Kamp68} or
scattering problems \cite{Grah02}. Casimir energies are also
relevant in the attempt of building a quantum theory of gravity
and of the universe \cite{Isha05}.

For these reasons, in Ref. \cite{Call02} we evaluated the force
produced by a weak gravitational field on a rigid Casimir cavity.
Interestingly, the resulting force was found to have opposite
direction with respect to the gravitational acceleration;
moreover, we found that the current experimental sensitivity of
small force macroscopic detectors would make it possible, at least
in principle, to measure such an effect \cite{Call02}.  In Ref.
\cite{Call02}, calculations were based on simple assumptions and
the result can be viewed as a reasonable ``{\it first order}''
generalization of $T_{\mu\nu}$ from Minkowski to curved
space-time. The present paper is devoted to a deeper understanding
and to more systematic calculations of the interaction of a weak
gravitational field with a Casimir cavity. To first order in our
approximation the former value of the force exerted by the field
on the cavity is recovered. 

We consider a plane-parallel Casimir cavity, made of ideal
metallic plates, at rest in the gravitational field of the earth,
with its plates lying in a horizontal plane. We evaluate the
influence of the gravity acceleration $g$ on the Casimir cavity
but neglect any variation of the gravity acceleration across the
cavity, and therefore we do not consider the influence of tidal
forces. The separation $a$ between the plates is taken to be much
smaller than the extension of the plates, so that edge effects can
be neglected. We obtain a perturbative expansion of the
energy-momentum tensor of the electromagnetic field inside the
cavity, in terms of the small parameter $\epsilon \equiv 2
ga/c^2$, to first order in $\epsilon$. For this purpose, we use a
Fermi \cite{Misn73}, \cite{Marz94} coordinates system $(t,x,y,z)$
rigidly connected to the cavity. The construction of these
coordinates involves only invariant quantities such as the
observer's proper time, geodesic distances from the world-line,
and components of tensors with respect to a tetrad \cite{Marz94}.
This feature makes it possible to obtain a clear identification of
the various terms occurring in the metric. In our analysis we
adopt the covariant point-splitting procedure \cite{Chri76},
\cite{Dewi75} to compute the perturbative expansion of the
relevant Green functions. Gauge invariance plays a crucial role
and we check it up to first order by means of the Ward identity.

With our notation, the $z$-axis coincides with the vertical
upwards direction, while the $(x,y)$ coordinates span the plates,
whose equations are $z=0$ and $z=a$, respectively. The resulting
line element for a non-rotating system is therefore \cite{Misn73}
$$
ds^{2}= -c^{2} \left(1+\epsilon {z\over a} \right) dt^{2}
+dx^{2}+dy^{2}+dz^{2} + {\rm O}(|x|^{2}) =
\eta_{\mu\nu}dx^{\mu}dx^{\nu} - \epsilon \frac{z}{ a} c^{2}
dt^{2},
$$
where $\eta_{\mu \nu}$ is the flat Minkowski metric ${\rm
diag}(-1,1,1,1)$.

\section{The Energy Momentum Tensor and the Point-Splitting
Procedure}

In the Point-Splitting procedure the Energy-Momentum Tensor
\begin{equation*} T^{\mu \nu} \equiv {2\over \sqrt{-g}}{\delta S
\over \delta g_{\mu \nu}} \nonumber\label{(2.7)}
\end{equation*}
is obtained by introducing an auxiliary quantity $\langle
T^{\mu\nu'}(x,x') \rangle $ which involves the action of a
differential operator on the Hadamard function \cite{Chri76},
\cite{Dewi75}. In the coincidence limit
\begin{equation*}
{\langle T^{\mu\nu}(x)\rangle=\lim_{x'\rightarrow x}\langle
T^{\mu\nu'}(x,x')\rangle},
\end{equation*} 
$\langle T^{\mu\nu}(x)\rangle$ 
is worked out. For QED (we use the Lorenz gauge \cite{Lore67}
to obtain the standard wave operator on the potential)
\begin{equation*} S[A_{\mu},\chi,\psi]=\int \left[-{1\over
4}F_{\mu \nu}F^{\mu \nu} -{1\over
2}(\nabla^{\mu}A_{\mu})^{2}+\chi^{; \alpha}\psi_{; \alpha}
\right]\sqrt{-g}~d^{4}x, \label{(2.2)}\nonumber
\end{equation*}
\bigskip
one gets
\begin{equation}
\langle T^{\mu \nu} \rangle = \langle T_{A}^{\mu \nu} \rangle
+\langle T_{B}^{\mu \nu}\rangle+\langle T_{\rm gh}^{\mu
\nu}\rangle, \label{(2.8)}\nonumber
\end{equation}
with
\begin{eqnarray} \langle F_{\rho\alpha}F_{\tau\beta}\rangle &=&
\lim_{x'\rightarrow x} \frac{1}{4}\left[H_{\alpha\beta';\rho\tau'}
+ H_{\beta\alpha';\tau\rho'}-H_{\alpha\tau';\rho\beta'}-
H_{\tau\alpha';\beta\rho'}
\right. \nonumber \\
& & \left.- H_{\rho\beta';\alpha\tau'}-
H_{\beta\rho';\tau\alpha'}+H_{\rho\tau';\alpha\beta'}+
H_{\tau\rho';\beta\alpha'}\right], \label{(2.11)}\\
\langle T_{A}^{\mu\nu} \rangle  &=& \lim_{x'\rightarrow x}
\left[-\frac{1}{4}\left(g^{\mu\rho}g^{\nu\tau}
-\frac{1}{4}g^{\mu\nu}g^{\tau\rho} \right) g^{\alpha\beta} \langle
F_{\rho\alpha}F_{\tau\beta}\rangle \right] \label{(2.12)} \\
\langle T_{B}^{\mu\nu} \rangle &=& \lim_{x'\rightarrow x} \Biggl[
-\frac{1}{4}g^{\alpha\beta}\left(
g^{\mu\rho}g^{\nu\tau}+g^{\mu\tau}g^{\nu\rho}-g^{\mu\nu}g^{\tau\rho}\right)
 \left(H_{\beta\tau';\alpha\rho}+
H_{\tau\beta';\rho\alpha'}\right)  \nonumber \\
& & + \frac{1}{8}g^{\alpha\beta}g^{\mu\nu}g^{\rho\tau}
\left(H_{\beta\tau';\alpha\rho'}+
H_{\tau\beta';\rho\alpha'}\right)\Biggr]\label{(2.13)} \\
\langle T_{\rm gh}^{\mu\nu}\rangle &=& \lim_{x'\rightarrow x}
\Biggl[ -\frac{1}{4}\left(
g^{\mu\alpha}g^{\nu\beta}+g^{\mu\beta}g^{\nu\alpha}
-g^{\mu\nu}g^{\alpha\beta}\right) \left(H_{;\alpha\beta'} +
H_{;\beta\alpha'} \right)\Biggr]
\end{eqnarray}
having defined
\begin{eqnarray*}
H_{\mu \nu}(x,x') & \equiv & \langle [A_{\mu}(x),A_{\nu}(x')]_{+}
\rangle\equiv H_{\mu \nu'},\\
H(x,x') & \equiv & \langle [\chi(x),\psi(x')]_{+} \rangle ,
\end{eqnarray*}
\be[A_{\alpha;\rho},A_{\beta;\tau} ]_{+}   \equiv
\lim_{x'\rightarrow x} \frac{1}{2}\left\{
[A_{\alpha';\rho'},A_{\beta;\tau}]_{+}
+[A_{\alpha;\rho},A_{\beta';\tau'}]_{+} \right\}.\ee

Since we need a recursive algorithm for the evaluation of Green
functions, it is more convenient to work with the Feynman Green
function instead of the Hadamard Green function. They are related
through
\begin{equation*}
H(x,x')=-2i[G(x,x')-\overline{G}(x,x')]
\end{equation*}
where $\overline{G}(x,x')=\frac{1}{2}[G^++G^-]$. The photon Green
function $G_{\lambda \nu'}$ in a curved spacetime with metric
$g_{\mu \nu}$ solves an equation of the form \cite{Bimo04} ($g_{\mu \nu'}$
being the parallel displacement bivector) \be \sqrt{-g}P^{\;
\lambda}_\mu(x) G_{\lambda\nu'}=g_{\mu\nu'}
\delta(x,x')=g_{\mu\nu} \delta(x,x'). \ee On expanding (this is,
in general, only an asymptotic expansion) \be G_{\lambda\nu'} \sim
G^{(0)}_{\lambda\nu'}+\epsilon \; G^{(1)}_{\lambda\nu'} + {\rm
O}(\epsilon^{2}), \ee we get, to first order in $\epsilon$,
\begin{eqnarray}
\Box^{0} G^{(0)}_{\mu\nu'}  &=& J^{(0)}_{\mu\nu'}, \\
\Box^{0} G^{(1)}_{\mu\nu'} &=& J^{(1)}_{\mu\nu'},
\end{eqnarray}
where
\begin{eqnarray*}
J^{(0)}_{\mu\nu'}  &\equiv&  -\eta_{\mu\nu}\delta(x,x'), \\
\epsilon J^{(1)}_{\mu\nu'} & \equiv & {z\over a} \epsilon \left
({\eta_{\mu\nu} \over
2}+\delta^0_{\mu}\delta^0_{\nu}\right)\delta(x,x') +2
\eta^{\rho\sigma}\Gamma^\tau_{\sigma\mu} G^{(0)}_{\tau\nu',\rho} +
\eta^{\rho\sigma}\Gamma^\tau_{\rho\sigma} G^{(0)}_{\mu\nu',\tau}
-{z\over a} \epsilon G^{(0)}_{\mu\nu',00}
\end{eqnarray*}
with $\Box^0\equiv\eta^{\alpha\beta}
\partial_\alpha\partial_\beta=
-\partial_0^2+\partial_x^2+\partial_y^2+\partial_z^2$.

To fix the boundary conditions we note that, on denoting by ${\vec
E}_{t}$ and ${\vec H}_{n}$ the tangential and normal components of
the electric and magnetic fields, respectively, a sufficient
condition to obtain
$$
\left . \vec{E}_{t} \right |_{S}=0,~~ \left . \vec{H}_{n} \right
|_{S}=0,
$$
on the boundary $S$ of the device, is to impose Dirichlet boundary
conditions on \cite{Espo97}
$$
A_0(\vec{x}),A_1(\vec{x}),A_2(\vec{x})
$$
at the boundary $z=0$, $z=a$. The boundary condition on $A_3$ is
determined by requiring that the gauge-fixing functional, here
chosen to be of the Lorenz type, should vanish on the boundary
(the boundary conditions on all components of $A_\mu$ are then all
preserved under gauge transformations \cite{Espo97} 
provided the same boundary
condition on ghost fields is imposed, i.e. homogeneous Dirichlet).
This implies
$$
\left . A^\mu_{;\mu} \right |_{S}= 0\Rightarrow \left . A^3_{;3}
\right |_{S} = \left . (g^{33}\partial_3
A_3-g^{\mu\nu}\Gamma^3_{\mu\nu}A_{3} )\right |_{S}=0 ~.
$$
To first order in $\epsilon$, these conditions imply for Green
functions the following:
\begin{eqnarray*}
\left . G^{(0)}_{\mu\nu'} \right|_{S} &=& 0, ~~ \left . \partial_3
G^{(0)}_{3\nu'} \right |_{S} ~=~ 0,~~\mu=0,1,2, \;
\forall \nu', \\
\left . G^{(1)}_{\mu\nu'} \right |_{S} &=& 0, ~~ \left .
\partial_3 G^{(1)}_{3\nu'} \right |_{S} ~=~ -{1 \over 2a} \left .
G^{(0)}_{3\nu'} \right |_{S},~~\mu =0,1,2, \; \forall \nu',
\end{eqnarray*}
hence we find that the third component of the potential $A_\mu$
satisfies homogeneous Neumann boundary conditions to zeroth order
in $\epsilon$ and inhomogeneous boundary conditions to first
order.

Now we are in a position to evaluate, at least formally (see
below), the solutions to zeroth and first order, and we get
\begin{equation}
G^{(0)}_{\lambda\nu'} = \eta_{\lambda\nu'} \int{ {d\omega d^2k
\over (2\pi)^3} e^{-i\omega(t-t')+ i {\vec k}_{\perp}\cdot({\vec
x}_{\perp}-{\vec x}_{\perp}')} } 
\Bigr[(1-\delta_{\lambda 3})g_{D}(z,z')+\delta_{\lambda 3}g_{N}(z,z')\Bigr],
\end{equation}
having defined
\begin{eqnarray*}
g_{D}(z,z';\kappa) &\equiv& {\sin{\kappa( z_<)}\sin{\kappa(a-z_>)}
\over \kappa\sin{\kappa a} },~~~~~~~0<z,z'<a, \\
g_{N}(z,z';\kappa) &\equiv& - {\cos{\kappa(
z_<)}\cos{\kappa(a-z_>)} \over \kappa\sin{\kappa a}
},~~~~0<z,z'<a,
\end{eqnarray*}
where $D,N$ stand for homogeneous Dirichlet or Neumann boundary
conditions, respectively, $z_>~(z_<)$ are the larger (smaller)
between $z$ and $z'$, while ${\vec k}_{\perp}$ has components
$(k_x,k_y)$, ${\vec x}_{\perp}$ has components $(x,y)$, $\kappa
\equiv \sqrt{\omega^2-k^2}$, and
\begin{equation} G^{(1)}_{\mu
\nu'}=\int{ {d\omega d^2k \over (2\pi)^3} e^{-i\omega(t-t')+i{\vec
k}_{\perp}\cdot({\vec x}_{\perp} -{\vec x}_{\perp}')} \Phi_{\mu
\nu'} },
\end{equation}
where the $\Phi$ components different from zero are written in
Ref. \cite{Bimo06}. The ghost field satisfies the same equations
as the $22$ component of the gauge field, hence we do not write it
explicitly. In the following we will write simply $G_{\mu\nu'}$
and $G$ for the Green function of the gauge and ghost field,
respectively.

We should stress at this stage that, in general, the integrals
defining the Green functions are divergent. They are well defined
as long as $x\neq x'$, hence we will perform all our calculations
maintaining the points separated and only in the very end shall we
take the coincidence limit as $x' \rightarrow x$ \cite{Endo84}. We
have decided to write the divergent terms explicitly so as to bear
them in mind and remove them only in the final calculations by
hand, instead of making the subtraction at an earlier stage.

Our Green functions are found to satisfy the Ward identity \be
G^\mu_{\; \nu';\mu}+G_{;\nu'}=0,~~~ G^{\mu \; \; ;\nu'}_{\;
\nu'}+G^{;\mu}=0, \ee to first order in $\epsilon$ so that, to
this order, gauge invariance is explicitly preserved. Ward
identities imply $ \langle T_{B}^{\mu \nu}\rangle+\langle T_{\rm
gh}^{\mu \nu}\rangle=0 $ to first order in $\epsilon$, thus in the
following we do not consider them. {\em Nonetheless we explicitly
computed them and verified that they cancel each other}.

\section{Energy-Momentum Tensor}

Using eqs. (1)-(3) we get, from the asymptotic expansion
$T_{\mu\nu'} \sim T^{(0)}_{\mu\nu'}+{\epsilon \over a}
T^{(1)}_{\mu\nu'} +{\rm O}(\epsilon^{2})$,
\begin{eqnarray}
\langle T^{(0)\mu\nu'}\rangle &=& {1 \over 16\,a^4\,{\pi }^2}
\left( {\zeta}_{H}\left(4, {2\,a + z - {z'} \over 2\,a}\right) +
{\zeta}_{H}\left(4, {z'-z \over 2\,a}\right) \right) \nonumber \\
& & \times{\rm diag}(-1,1,1,-3),
\end{eqnarray}
where $\zeta_{H}$ is the Hurwitz $\zeta$-function
$\zeta_{H}(x,\beta) \equiv \sum_{n=0}^{\infty}(n+\beta)^{-x}$. On
taking the limit $z'\rightarrow z^+$ we find
\begin{equation}
\lim_{z' \to z^+} \langle T^{(0)\mu\nu'} \rangle =\left({\pi^2
\over 720 a^4} +\lim_{z' \to z^+} {1 \over \pi^2(z-z')^4}\right)
{\rm diag}(-1,1,1,-3),
\end{equation}
where the divergent term as $z' \rightarrow z$ can be removed by
subtracting the contribution of infinite space without bounding
surfaces \cite{Bord01}, and in our analysis we therefore discard
it hereafter. The renormalization of the energy-momentum tensor in
curved spacetime is usually performed by subtracting the $\langle
T_{\mu \nu} \rangle$ constructed with an Hadamard or
Schwinger--DeWitt two-point function up to the fourth adiabatic
order \cite{Chri76}, \cite{Chri78}. In our problem, however, as we
work to first order in $\epsilon$, we are neglecting tidal forces
and therefore the geometry of spacetime in between the plates is
flat. Thus, we need only subtract the contribution to the energy
momentum tensor that is independent of $a$, which is the standard
subtraction in the context of the Casimir effect in flat
spacetime \cite{Deut79}.

In the same way we get, to first order in $\epsilon$:
\begin{eqnarray}
 \lim_{z' \to z^+} \langle
T^{(1)\mu\nu'}\rangle &=& {\rm
diag}(T^{(1)00},T^{(1)11},T^{(1)22},T^{(1)33}) \nonumber \\
& +& \lim_{z' \to z^+} {\rm diag}\Bigr(-z'/\pi^{2}(z-z')^{4},0,0,0
\Bigr),
\end{eqnarray}
where
\begin{eqnarray}
T^{(1)00} &=& -{{\pi }^2 \over 1200\,a^3} + {11 {\pi }^2\,z \over
3600\,a^4} - {\pi  \over 60\,a^3}\, \frac{ \cos{({\pi \,z \over
a})}}{{\sin^{3}{ ({\pi \,z \over a})}}}, \\
T^{(1)11} &=& {{\pi }^2 \over 3600\,a^3} - {{\pi}^2\,z \over
1800\,a^4} -{\pi \over 120\,a^3}\, \frac{ \cos{({\pi \,z \over
a})}}{{\sin^{3}{ ({\pi \,z
\over a})}}}, \\
T^{(1)22} &=& T^{(1)11}, \\
T^{(1)33} &=& -{\left( {\pi }^2\,\left( a - 2\,z \right) \right)
\over 720\,a^4}.
\end{eqnarray}
Incidentally we note that the tensor is covariantly conserved:
$\nabla\cdot T=0$ to first order in $\epsilon$.

\section{Push}

To compute the Casimir energy we must project the energy-momentum
tensor along the unit timelike vector $u$ with covariant
components $u_\mu=(\sqrt{-g_{00}},0,0,0)$ to obtain $\rho=\langle
T^{\mu\nu}\rangle u_\mu u_\nu$, so that 
\be 
\rho = -{\pi^2 \over
720a^4}+2 {g \over c^2} \left(-{{\pi }^2 \over 1200\,a^3} + {{\pi
}^2\,z \over 600\,a^4} - {\pi \over 60\,a^3}\, \frac{ \cos{({\pi
\,z \over a})}}{{\sin^{3}{ ({\pi \,z \over a})}}} \right) 
+{\rm O}(g^{2}), 
\ee 
where we have replaced
$\epsilon$ by its expression in terms of $g$. Thus, the energy
stored in the Casimir device is found to be 
\be
E=\int{d^3\Sigma\sqrt{-g}\langle T^{\mu\nu}\rangle u_\mu
u_\nu}=-{\hbar c \pi^2 \over 720} {A \over a^{3}} \left(1+{1 \over
2} {g a \over c^2} \right)\equiv E_C\left(1+{1 \over 2} {g a \over
c^2} \right), 
\ee 
where $A$ is the area of the plates, $d^3\Sigma$
is the three-volume element of an observer with four-velocity
$u_\mu$, and we have reintroduced $\hbar$ and $c$.

In the same way, the pressure on the plates is given by 
\be
P(z=0)={\pi^{2}\over 240} {\hbar c \over a^{4}} \left(1+{2\over
3}{ga \over c^{2}}\right), \; P(z=a)=-{\pi^{2}\over 240 } {\hbar c
\over a^{4}} \left(1-{2\over 3} {ga \over c^{2}}\right). 
\ee 
To obtain the resulting force one has to multiply each of them by the
redshift $r$ of the point where they act, relative to the point
where they are added \cite{nordt75}:
 \be 
r_{\tiny P_{added}(\tiny
P_{act})}=\sqrt{\frac{|g_{00}(P_{act})|}{|g_{00}(P_{added})|}}
\approx 1+\frac{g}{c^2}(z-z_Q)
\ee 
to leading order in $\frac{g
z}{c^2}$, so that a net force \cite{bica071} 
\be 
F =-\frac{\pi^2
\hbar c}{a^4}\left[ \frac{g}{240 c^2}(z_2-z_1)-
 \frac{4 g}{720 c^2}(z_2-z_1)\right]
 = {\pi^{2}\over 720}{A \hbar g\over c a^{3}}=\frac{E_C}{c^2}g,
\ee 
pointing upwards along the $z$-axis is obtained, in perfect
agreement with the early findings of \cite{jare93} and the more
recent results of \cite{fumi07}
(for other relevant references on curved spacetime calculations,
see \cite{Endo84, Chri78, Dewi84}).

\section{Concluding Remarks}

To the best of our knowledge, the analysis presented in this paper
represents the first study of the energy-momentum tensor for the
electromagnetic field in a Casimir cavity placed in a weak
gravitational field. The resulting calculations are considerably
harder than in the case of scalar fields. By using Green-function
techniques, we have evaluated the influence of the gravity
acceleration on the regularized energy-momentum tensor of the
quantized electromagnetic field between two plane-parallel ideal
metallic plates, at rest in the gravitational field of the earth,
and lying in a horizontal plane. In particular, we have obtained a
detailed derivation of the theoretical prediction according to
which a Casimir device in a weak gravitational field will
experience a tiny push in the upwards direction \cite{Call02}.
This result is consistent with the picture that the {\it negative}
Casimir energy in a gravitational field will behave like a {\it
negative mass}. An outstanding open problem is now how
to obtain an independent evaluation of our formula for the
energy-momentum tensor, and what its implications are for fundamental
physics.

\section*{Acknowledgments}
The work of G. Bimonte and G. Esposito has been partially
supported by PRIN {\it SINTESI}. G. Esposito is grateful to the
Dipartimento di Scienze Fisiche of Federico II University, Naples, 
for hospitality and support. The work of L. Rosa has been
partially supported by PRIN {\it FISICA ASTROPARTICELLARE}.

\end{document}